\documentstyle[aps,prl,multicol,epsf]{revtex}
\def\asigma{\stackrel{\leftrightarrow}{\sigma}}
\def\aU{\stackrel{\leftrightarrow}{U}}

\def\gdot{\dot{\gamma}}

\def\gs{\gtrsim}
\def\ls{\lesssim}
\def\be{\begin{equation}}
\def\en{\end{equation}}                  
\def\p{\partial} 
\newcommand{\bi}[1]{\mbox{\boldmath$#1$}}
\newcommand{\av}[1]{\langle{#1}\rangle}

\def\bea{\begin{eqnarray}}
\def\ena{\end{eqnarray}}

\renewcommand{\theequation}{\arabic{section}.\arabic{equation}}

\begin{document}
\draft
\bibliographystyle{prsty}
\title{ Plastic Flow in Two-Dimensional Solids}
\author{ Akira  Onuki}
\address{Department of Physics, Kyoto University, Kyoto 606-8502}
\date{\today}
\maketitle

\begin{abstract} 
A  time-dependent Ginzburg-Landau model  of 
plastic deformation in two-dimensional solids 
is presented.  The fundamental dynamic variables 
are the displacement field $\bi u$ and 
the  lattice velocity ${\bi v}=\p {\bi u}/\p t$. 
Damping is assumed to arise from the  
shear viscosity in the momentum 
equation. The elastic energy density is a 
periodic function of  the 
shear and tetragonal strains, 
which enables    formation of slips at large strains.  
In this work  
we neglect defects such as 
 vacancies, interstitials, or  grain boundaries. 
The simplest slip consists  
of two edge dislocations with opposite Burgers vectors.
The  formation energy of  a slip is  minimized  
if  its  orientation  is parallel or perpendicular 
to the flow in simple shear deformation 
and  if it makes  angles of $\pm \pi/4$ 
with respect to the stretched direction in uniaxial 
stretching.   High-density dislocations produced    
in plastic flow do not disappear even if 
the flow is stopped. Thus  large applied strains 
give  rise to metastable,  structurally disordered  states. 
We divide the  elastic energy into 
an elastic part due to affine deformation 
and a defect part. The latter  represents 
 degree of   disorder and is   nearly 
 constant in plastic flow under  cyclic straining.@

\end{abstract}
\pacs{PACS numbers: 62.20.Fe, 61.72.Lk, 81.40.Lm, 83.20.Jp}

\begin{multicols}{2}

\date{ }
\pagestyle{empty}

\section{Introduction}

Plastic flow  
has long been studied 
in crystalline and amorphous 
 solids and in glassy polymers.  In crystals 
irreversible motions of dislocations 
give rise to plastic deformation 
and large strains produce 
high-density dislocations 
\cite{Friedel,Nabarro}. 
The nonlinear flow properties 
are very sensitive to the amount of such defects 
and  strongly dependent on the deformation history. 
Simulations of dynamics of dislocation lines 
have recently been performed 
 but are still most  difficult 
\cite{Kubin,Ab,Schwarz}.

In amorphous solids   at low temperature $T$ 
\cite{Gilman,Spaepen,Argon,Spaepen_review,Kimura,Spaepen_review1,Argon_review}, salient features are asfollows. 
(i)  Shear strains tend to be localized 
in narrow shear bands in plastic flow above 
a yield stress. The  width of 
such shear bands is microscopic  in the initial stage 
\cite{band_micro} 
but can grow to mesoscopic  sizes \cite{band}, 
sometimes resulting in fracture.  Shear bands 
were numerically realized at large shear strains 
in  molecular dynamics (MD) 
simulations of  two-dimensional (2D) two-component 
glasses \cite{Deng,Falk} and in  simulations  
of a 2D  phenomenological stochastic model \cite{Bulatov}.  
(ii) As another aspect, 
in  3D MD simulations  on model two-component 
 glasses  at low $T$,  
Takeuchi {\it et al.} \cite{Takeuchi}
 observed heterogeneities among 
mobile and immobile regions 
after application of  shear strains.   In   2D and 3D MD 
simulations  on model supercooled binary mixtures 
  above $T_{\rm g}$,  
similar  or much more extended 
dynamic heterogeneities 
 have  been  detected   
 in quiescent states  
 \cite{Muranaka,Harrowell,Yamamoto1,Donati} 
and found to be sensitively suppressed by 
 applied shear flow \cite{Yamamoto}. 
(iii) Furthermore, at higher $T (\gs 0.7-0.8 T_{\rm g})$ 
(where $T_{\rm g}$ is the glass transition temperature), 
 shear deformation  occurs 
 quasi-homogeneously 
(still involving many particles 
in each configuration change),  
leading to highly viscous non-Newtonian 
 behavior \cite{Yamamoto,YamamotoJ,Simons,Chen}.

Glassy polymers also behave analogously \cite{Argon_review,Crist,Boyce},
 where  shear bands appear 
above a yield stress  at low $T$ 
\cite{Argon_p},  and   highly viscous  non-Newtonian 
flow and significant elongation  of the chain shapes occur 
at elevated  $T$ \cite{YamamotoJ,YamamotoJCP}.  
In glassy polymers,  
the entropic stress arising from 
 molecular orientations becomes significant 
at large strains and   such systems behave like 
 cross-linked rubbers 
\cite{Argon_review,Crist,Boyce}.

Recently much  attention has 
been paid to {\it jamming  rheology} 
observed in sheared states 
of  supercooled liquids, 
soft glassy materials  such as 
dense microemulsions, or granular materials 
\cite{Liu_book}.  In  these systems, the thermal agitation effect 
is very small if the  particle size is large, 
but universal constrained dynamics is realized 
 under external forces. In supercooled liquids (at relatively 
high $T$) \cite{Yamamoto} and dense 
microemulsions  (at effectively 
low $T$) \cite{Okuzono,Durian}, 
 mesoscopic dynamic heterogeneity and strong shear-thinning 
behavior have been observed,  but  shear bands have not 
been identified. In granular materials (at effectively zero temperature), 
strain localization is most conspicuous 
\cite{granular,granular1,Behringer}.

In our recent work \cite{OnukiJ} 
we have constructed  2D nonlinear strain 
theory taking into account 
 the underlying local periodic lattice structure, 
 where the elastic energy density 
is periodic with respect to the shear and tetragonal strains.  
  Then in  our model 
plastic flow starts with  appearance of slips.  
The simplest  slips consist of two edge dislocations having 
opposite Burgers vectors  with size $a$ ($a$ being the lattice 
constant) and they  
  grow into mesoscopic 
shear bands as the applied strain is increased. 
Under uniaxial stretching \cite{uni},   
well  developed shear bands were already 
 numerically realized \cite{Falk,Bulatov} 
and will be realized also in our  simulations.  
These   shear bands make 
  angles of   $\pm \pi/4$  
with respect to the stretched direction in 
agreement   with  observations 
 in various amorphous materials 
\cite{Spaepen,Argon,Spaepen_review,Kimura,Crist}   
 including granular materials \cite{granular,granular1}.  
 These angles   will be  shown 
to minimize the elastic energy of 
incipient  slips  in this paper.

In  crystalline  solids,  dislocation pairs 
in 2D or dislocation loops in 3D forming slip lines or surfaces 
should be nucleated  at the inception  of 
plastic deformation (in addition 
to preexisting dislocations). 
In   amorphous solids,  
it has been  controversial on  whether dislocations
 themselves can be well-defined or not \cite{Gilman,Spaepen_review}.  
It is not obvious how to characterize 
the local rearrangement processes 
 as to their shapes and sizes.  
In MD simulations on binary mixtures,  
they  have been 
detected as cluster-like objects 
 with various visualization methods
\cite{Takeuchi,Muranaka,Harrowell,Yamamoto1,Donati,Yamamoto,YamamotoJ,Langer}. 
If the size-ratio of the constituent two species is 
chosen such that crystallization is  most suppressed, 
strong frustration occurs in the packing of large and small 
particles in jammed states.  Then the local crystalline order 
can be well-defined only over  short distances. 
However, in  a 2D amorphous 
soap bubble raft,   Argon and Kuo   \cite{Kuo} observed   that 
nucleation of a  dislocation pair 
gave  rise to a small-scale slip  
but  such dislocations  did  not 
glide more than 2-3 bubble distances. 
It is worth noting that 
Deng {\it et al.} \cite{Deng} 
 found extended slip-like strain localization 
in 2D MD simulations with the size-ratio 
rather close to 1 (see comment (iv) in the last section).  
Notice that the displacement field around  a slip 
is localized  because 
the two constituent dislocations have  
opposite Burgers vectors.  
As a result, small-scale slips 
 should be well-defined 
even in amorphous  solids 
 as long as  
the slip size does not much exceed  the 
range of the  local  crystal structure.

The plastic flow phenomena are thus very complex, 
being influenced by many factors, 
but they are ubiquitous  in various  kinds of solid-like  materials.  
The   purpose of this paper is  
 to present a well-defined  Ginzburg-Landau model consistently taking 
account of nonlinear elasticity.   A merit of this approach  
is that we can  put emphasis on any  aspects of the phenomena 
by  controlling   the parameters or changing  
the model itself. We will  examine 
(i) the fundamental flow units, slips, 
in  detail numerically and analytically 
 and (ii)  plastic flow numerically in 
simple shear and elongational (stretching) deformation. 
To make this paper simplest,  
as  it is the  first detailed exposition 
of our scheme,   we will   neglect 
(i)   vacancies and interstitials,  or 
a variable $m$ representing the local free-volume. 
A   dynamic model including 
such an additional degree of freedom 
has already   been presented in our previous work \cite{OnukiJ}.
We will also neglect (ii) the configurational frustration 
effect induced by the size difference between the two species. 
Introducing these two ingredients 
 will constiture future development of our scheme.

This paper is organized as follows. 
In Section 2 we will present 
our dynamic model and explain 
 our numerical method.
In Section 3 we will discuss 
the simplest form of slips 
numerically obtained from our 
nonlinear elasticity theory.  We will  
also derive  some  analytic expressions 
for the slip formation energy under 
general strain field starting with 
the Peach-Koehler theory \cite{Peach}  
and  compare them with 
 numerical results. 
In our  scheme   stationary  slip solutions 
exist for small externally applied strains, 
where the force balance is achieved in the presence of  
the Peierls potential energy 
for  the dislocation position in crystals 
\cite{P,N}.    Section 4  will present 
numerical results of the stress-strain relations  
and the patterns of the strains, 
 the elastic energy density, and the displacement 
vector  under applied strains. We will 
also give a method of dividing the elastic 
energy into an elastic part 
due to affine deformation and a defect part.

\setcounter{equation}{0}
\section{Model equations}
\subsection{Elastic energy}

Recently Doi {\it et al.} \cite{Doi}
described plastic flow 
in a highly viscous 
2D crystal phase of block copolymers  
in the framework of  a 1D Frenkel-Kontrowa model \cite{Lubensky} 
assuming  1D sliding motions.  
Hereafter we present  a  2D  Frenkel-Kontrowa model 
to describe plastic flow in 2D 
\cite{OnukiJ}. In terms of  the displacement 
vector ${\bi u}=(u_x,u_y)$ from a reference crystal state, 
we   define the 
strain components as 
\bea 
e_1&=&\nabla_x u_x+ 
\nabla_yu_y , \nonumber\\
e_2&=& \nabla_x u_x- 
\nabla_yu_y, \nonumber\\
e_3&=& \nabla_x u_y 
+\nabla_yu_x, 
\label{eq:2.1}
\ena 
where $\nabla_x=\p/\p x$ and 
$\nabla_y=\p/\p y$.  We call $e_1$ the dilation strain, 
$e_2$ the tetragonal strain, and $e_3$ the shear strain. 
If we suppose a 2D triangular lattice with lattice constant $a$,  
the elastic energy should be invariant with 
respect to the rotations  of the reference frame 
by $\pm n\pi/3$ ($n=1,2, \cdots$). Due to this symmetry, 
the elasticity must be  isotropic in the harmonic 
approximation \cite{Onukibook}, being characterized  
by the bulk and shear moduli,  $K_0$ and $\mu_0$, 
but it depends on the orientational angle $\theta$ 
of one of the crystal axes  with respect to the $x$ axis 
for large shear strains. 
Under rotation of the reference frame 
by $\theta$, the shear strains $e_2$ and $e_3$ are changed 
to  $e_2'$ and $e_3'$, where \cite{Onukibook}
\bea 
e_2'&=& e_2 \cos 2\theta + e_3 \sin 2\theta, \nonumber\\ 
e_3'&=& e_3 \cos 2\theta -e_2 \sin 2\theta .  
\label{eq:2.2}
\ena 
These relations are obtained from 
the orthogonal transformations, 
${\bi r}'= \aU\cdot{\bi r}$ and 
${\bi u}'= \aU\cdot{\bi u}$, where 
${\bi r}'= (x',y')$ and  ${\bi u}'=  (u'_x,u'_y)$ 
represent    the position and the displacement, 
respectively,  in the new reference frame, 
and $\aU= \{U_{ij}\}$  with $U_{xx}=U_{yy}=\cos\theta$ 
and $U_{xy}=-U_{yx}=\sin\theta$.  
The $e_2'$ and $e_3'$ are the tetragonal and shear strains, respectively, 
in the new reference frame where the  $x'$ axis is along one of the 
crystal axes.

The  elastic energy is written as $F_{\rm el} = \int d{\bi r} f_{\rm el} $
with the elastic energy  density in  the form, 
\be 
f_{\rm el}= \frac{1}{2}{K_0}e_1^2 + \mu_0 {\Phi}(e_3', e_2'), 
\label{eq:2.3}
\en 
which   is  independent of the rotation strain,  
\be 
\omega=\nabla_x u_y 
-\nabla_yu_x.  
\label{eq:2.4}
\en 
Note that $e_1$ and $\omega$ are invariant with 
respect to the rotation of the reference frame. 
The simplest form of     ${\Phi}$ is given by 
\bea 
{ \Phi}(e_3', e_2') &=&  \frac{1}{6\pi^2} 
\bigg [ 3- \cos\pi(\sqrt{3}e_3'-e_2') \nonumber\\
&&\hspace{-2cm} - \cos\pi(\sqrt{3}e_3'+e_2')
-  \cos(2\pi e_2') \bigg ]. 
\label{eq:2.5}
\ena  
This function is invariant with respect to the rotation 
$\theta \rightarrow  \theta +  \pi/3$,  
is a periodic function of $e_3'$  
with period $2/\sqrt{3}$ for $e_2'=0$ 
(simple shear deformation),  and 
becomes $(e_2^2+e_3^2)/2$ for small strains.  
In Fig.1 we display $\Phi(e_3,e_2)$ 
supposing $\theta=0$, where 
 one of the crystal axes is along the $x$ axis. 
We can see a hexagonal lattice 
structure in the $e_3$-$e_2$ plane. 
 As a characteristic feature,  Fig.2 shows that it is almost isotropic 
or is a function of $e= (e_3^2+e_2^2)^{1/2}$ only even for 
rather large $e$ ($\ls 0.5$) (in the 
unit cell at the origin).  In fact, we have the Taylor expansion, 
\bea 
\Phi(e_3,e_2)&=& \frac{1}{2}e^2-\frac{\pi^2}{8}e^4
+ \frac{\pi^4}{72}e^6  \nonumber\\
&&\hspace{-1.5cm} + \frac{\pi^4}{720} 
(e_2^2-e_3^2)(e^4-16e_2^2e_3^2)  
+ O(e^8) 
\label{eq:2.6}
\ena 
If we set $e_3 = e \cos\chi$ and 
$e_2 = e \sin\chi$, the fourth  term is rewritten as 
$-({\pi^4}/{720}) e^6 \cos (6\chi)$ and 
is known to keep the invariance  
$\chi \rightarrow \chi+\pi/3$. 
Anisotropy appears 
in the terms of order $e^6$, but 
 the anisotropic fourth term is at most $10\%$ of
 the isotropic third  term.

We  examine   elastic  
stability  of homogeneously 
strained states. 
That is, we  superimpose  
 infinitesimal  strains, $\delta e_3$ and $\delta e_2$,  
 on  $e_3$ and $e_2$ assumed to be homogeneous. 
The  second order terms  in 
$\Phi(e_3+\delta e_3,e_2+\delta e_2)$
read 
\be 
\delta^{(2)}\Phi= \frac{\Phi_{33}}{2}(\delta e_3)^2
+ \Phi_{23}\delta e_3\delta e_2 + 
\frac{\Phi_{22}}{2}(\delta e_2)^2,  
\label{eq:2.7}
\en 
where $\Phi_{\alpha\beta}=
\p^2\Phi/\p e_\alpha \p e_\beta$ ($\alpha, \beta=2,3$). 
In the stable regions the above second 
order contribution should be positive-definite 
 or the two eigenvalues, $\lambda_1$ and $\lambda_2$, 
 of the $2\times 2$ matrix 
$\{\Phi_{\alpha\beta}\}$ should be both positive. 
In the appendix we will derive this linear stability 
criterion by solving the linearized  version 
of our dynamic model to be presented in the next section. 
Fig.3 shows  that the system is stable for $e \ls 0.3$, where 
$\Phi$ depends almost only on $e$. 
This result readily follows if we neglect the anisotropy in 
$\Phi$ by setting $\Phi(e_2,e_3) \cong  G(e^2)$. Then some calculations 
yield  
\be 
\Phi_{\alpha\beta}\cong  2G'\delta_{\alpha\beta}
 + 4G'' e_\alpha e_\beta, 
\label{eq:2.8}
\en 
where $G'= dG(e^2)/de^2$ and $G''=dG'(e^2)/de^2$. 
From (2.6) we find $G'\cong 1/2$ and $G'' \cong -\pi^2/4$. 
The determinant of $\{\Phi_{\alpha\beta}\}$ 
becomes $4G'(G' + 2G'' e^2)$. Thus 
the stability condition 
becomes 
\be 
e< (G'/|2G''|)^{1/2} \cong  0.3,
\label{eq:2.9}
\en  
around the origin in the $e_3$-$e_2$ plane. 
The elastic instability with a negative 
eigenvalue ($\lambda_1<0$ 
or  $\lambda_2<0$)  causes  rapid relaxation processes 
resulting in localized slips and stress release. In plastic flow 
  the  stability  condition is  mostly 
satisfied throughout the system (see Fig.11).

If we assume that small solid elements 
are rotated  without shape changes  with 
the local angular  velocity $(\p v_y/\p x- \p v_x/\p y)/2$, 
 the orientation angle $\theta$ 
is related to the rotation strain  as  
\be 
\theta = \frac{1}{2} \omega  +\theta_0 , 
\label{eq:2:10}
\en 
where $\theta_0$ is the initial value independent of 
time $t$.  This relation   was assumed 
in our previous paper \cite{OnukiJ}. 
However, it becomes not well defined 
when the three crystal axes are rotated  differently. 
For example, let  a  crystal 
with the three axes at $\theta=0, \pi/3,$ and $-\pi/3$ 
be affinely  deformed by  a simple  
shear deformation given by  
$u_x=\gamma y$ and $u_y=0$ ($e_3=\gamma$ and $e_2=0$). 
Then the  the first axis along the $x$ direction 
is not rotated, while the other axes are rotated by 
\be 
\delta\theta_{\pm}=\pm \bigg [ {\rm tan}^{-1} \bigg ( 
\frac{\sqrt{3}}{1\pm \sqrt{3}\gamma} \bigg )- \frac{\pi}{3}\bigg ].
\label{eq:2.11}\en  
For $\gamma \ll 1$, 
 we have  $\delta\theta_{\pm} \cong  -3\gamma/4$, 
so   (2.10)  holds   for the 
 average of the three rotation  angles, 
$(0+\delta\theta_+ +\delta\theta_-)/3=-\gamma/2=
\omega/2$. 
Thus there  remains  ambiguity in 
(2.10) particularly at large strains 
$e \gs 0.5$ \cite{commentp}. 
In this paper, in view of 
the virtual isotropic behavior 
 of our  elastic energy for $e \ls 0.5$, 
 we will simply set 
\be 
\theta =0,  
\label{eq:2.12}
\en 
throughout the system. 
To check the appropriateness 
of this assumption,   we 
have also performed simulations with   
various  homogeneous $\theta$ held fixed. 
The resultant 
patterns and stress-strain relations 
are  insensitive 
to $\theta$ and very  similar to those 
obtained  for 
$\theta=0$  to follow.

\subsection{Dynamic equations}

The elastic stress tensor should 
be  defined for general strains. 
Let us change an arbitrary displacement  $\bi u$ 
by  an infinitesimal 
amount $\delta{\bi u}$ as 
${\bi u} \rightarrow {\bi u}+\delta{\bi u}$. 
The incremental change of the elastic energy density is written as 
\cite{Onukibook} 
\be 
\delta f_{\rm el} = 
\sum_{ij} \sigma_{ij}\frac{\p}{\p x_j} \delta u_i.
\label{eq:2.13}
\en 
This is just the definition of the elastic stress  tensor 
$\sigma_{ij}$. Under (2.10) \cite{commentp} it is 
symmetric and its components 
are expressed as   
\bea 
\sigma_{xx} &=& K_0 e_1+  {\mu_0}A_2 ,\quad  
\sigma_{yy}= K_0 e_1 -  {\mu_0}A_2,\nonumber\\ 
\sigma_{xy}&=& \sigma_{yx}=  {\mu_0}A_3,
\label{eq:2.14}
\ena  
where  $A_\alpha = \p \Phi(e_3,e_2)/\p e_\alpha$ are written as 
\bea
A_2 &=& \frac{1}{{3}\pi} \bigg [ \cos (\sqrt{3} \pi e_3) 
\sin (\pi e_2) +\sin (2\pi e_2) \bigg ] , \nonumber\\
A_3 &=&  \frac{1}{\sqrt{3}\pi} \sin (\sqrt{3} \pi e_3) 
\cos(\pi e_2) .
\label{eq:2.15}
\ena 
For small strains we have  
$A_2 \cong e_2$ and $A_3 \cong e_3$ to reproduce  
 the isotropic, linear  elastic theory.

We assume that the lattice velocity 
\be 
{\bi v}= \frac{\p}{\p t}  {\bi u}
\label{eq:2.16}
\en 
obeys 
\be
\rho \frac{\p}{\p t}{\bi v} =  \nabla\cdot\asigma + 
 +\eta_0 \nabla^2v + \nabla \cdot \asigma_{\rm R}  ,    
\label{eq:2.17}
\en 
where   we introduce the shear viscosity $\eta_0$  
but neglect the bulk 
viscosity \cite{Landau,commentV}. 
The   $\asigma_{\rm R} =
\{\sigma^{\rm R}_{ij}\}$ is a symmetric random stress tensor 
and satisfies $\sigma^{\rm R}_{xx}+\sigma^{\rm R}_{yy}=0$, because the bulk 
viscosity is neglected, and is related to $\eta_0$ by \cite{Onukibook} 
\be 
\av{\sigma^{\rm R}_{ij}({\bi r},t)
\sigma^{\rm R}_{ij}({\bi r}',t')}=
2k_{\rm B}T \eta_0 \delta({\bi r}-{\bi r}')\delta(t-t').  
\label{eq:2.18}
\en 
The mass density  $\rho$ will be treated as  a constant in (2.17). 
This is justified when the deviation 
$\delta\rho=\rho -\av{\rho}$ is assumed to be 
 much smaller than the average $\av{\rho}$ \cite{commentc}. 
In the usual linear elasticity theory \cite{Landau} 
it is related to the dilation strain $e_1$ as 
$\delta\rho\cong  -\rho e_1$, so we are assuming $|e_1| \ll 1$ 
(and coincidence of the lattice and  
mass velocities \cite{OnukiJ}). 
In our simulation  
in the plastic flow regime at 
$\gdot=10^{-3}$, for example, 
$|e_1|$ attains  a maximum  in a range of  $0.2$-$0.3$ 
and the variance 
$\sqrt{\av{e_1^2}}$ 
increases up to about    $0.06$.

Due to the presence of the random stress,  
$\bi u$ and $\bi v$ are random variables 
and (2.16) and (2.17) constitute 
nonlinear  Langevin equations \cite{Onukibook}. 
Their   equilibrium distribution 
attained in the unstrained condition  is 
given by 
$P{\propto}\exp (-F/{k_{\rm B}T})$,  where 
the total free energy $F$ is the sum of the elastic energy 
$F_{\rm el} $ and the kinetic energy as  
\be 
F = \int d{\bi r}\bigg 
(f_{\rm el}  + \frac{\rho}{2}{\bi v}^2\bigg ) .
\label{eq:2.19}
\en 
Furthermore, if the random stress is 
omitted in (2.17) and the dynamic equations 
are treated as deterministic ones, 
the time derivative of $F$ 
is nonnegative-definite  in the unstrained condition as  
\be 
\frac{d}{dt}F = - \int d{\bi r} \sum_{ij}\eta_0 (\nabla_i v_j)^2 \le 0,  
\label{eq:2.20}
\en   
where  use is made of the relation, 
\be 
\nabla\cdot\asigma= -
\frac{\delta}{\delta {\bi u}} F.  
\label{eq:2.21}
\en 
The stationary condition 
  $d F/dt=0$ is attained for ${\bi v}={\bi 0}$ 
and $\nabla\cdot\asigma={\bi 0}$. 
This is a  condition to guarantee 
self-consistency of the dynamic equations 
which have stable equilibrium solutions.

In the appendix we will examine the linearized dynamic equations 
of our model around homogeneously strained states 
to obtain two sound modes in the stable region.

\subsection{Dimensionless forms}

We make our equations dimensionless 
 measuring   space  in units of the lattice constant $a$ and  
time in units of 
\be 
\tau_0= (\rho/\mu_0)^{1/2}a. 
\label{eq:2.22}
\en 
The stress components and 
the elastic energy density are 
measured in units of $\mu_0$ and the elastic energy  
in units of $\mu_0 a^2$, while the 
strains remain unscaled.  To avoid to introduce 
too many symbols, 
we will rewrite the scaled position vector $a^{-1}{\bi r}$, 
time $\tau_0^{-1}t$, displacement 
vector $a^{-1}{\bi u}$, and 
velocity $\tau_0a^{-1} {\bi v}$ 
simply as $\bi r$, $t$, $\bi u$, and $\bi v$ 
using  the original  notation.  Then, in the dimensionless $\asigma$,  
 $\mu_0$ is replaced by 1 and $K_0$   by the ratio, 
\be 
\lambda = K_0/\mu_0 .
\label{eq:2.23}
\en 
In terms of the  scaled quantities 
the equilibrium distribution 
is written as 
\be 
P\propto 
\exp  
\bigg [- \frac{1}{\epsilon_{\rm th}}
\int d{\bi r} \bigg (\frac{\lambda}{2} e_1^2
+ \Phi(e_3,e_2)+ \frac{1}{2}{\bi v}^2\bigg ) 
\bigg ], 
\label{eq:2.24}
\en 
where 
\be 
\epsilon_{\rm th} =  k_{\rm B}T /\mu_0 a^2 .
\label{eq:2.25}
\en  
The  parameter $\epsilon_{\rm th}$ represents the degree  of 
the thermal fluctuations (being proportional to $T$) 
and is an important 
parameter, for example,  in describing the decay of metastable 
states by thermal agitations. If the dynamic equation 
 (2.17) is  made dimensionless, the dimensionless viscosity is given by 
\be 
\eta_0^* = \eta_0a^{-1}(\rho\mu_0)^{-1/2} .
\label{eq:2.26}
\en  
In (2.18)  the noise strength 
$k_{\rm B}T\eta_0$  is replaced by 
$\epsilon_{\rm th} \eta_0^*$.

\subsection{Numerical method}

We integrate (2.16) and (2.17) in the dimensionless units 
on a $128\times 128$ square lattice represented by  
$(n,m)$ ($1 \leq n, m \leq 128$)  with $x= n\Delta x$ and 
$y=m \Delta x$.  The modulus ratio $\lambda$ in (2.23) is set equal to 1.
For simplicity, the  mesh size $\Delta x$ 
 and the dimensionless viscosity in 
(2.26)   are set equal to 1:
\be 
\Delta x= 1, \quad \eta_0^*=1.
\label{eq:2.27}
\en 
The first relation means that the mesh size is 
just equal to the lattice constant, 
and the second one is rewritten as  
$\eta_0= a (\rho\mu_0)^{1/2}$.

We give our   boundary conditions employed. 
At the bottom $y=0$, we set  ${\bi u}={\bi 0}$. 
At the top $y=L_0=128$,  
we set $u_x= \gamma L_0$ and 
$u_y=0$   in  the presence of 
applied shear strain $\gamma$, and 
we set $u_x= -u_y= \epsilon L_0/2$ in  the presence of 
applied tetragonal  strain $\epsilon$. 
In the $x$ direction, 
 we impose the periodic boundary condition, 
${\bi u}(x+L_0,y,t)= {\bi u}(x,y,t)$. 
The unstrained condition is realized for 
$\gamma=0$ and $\epsilon=0$.

We are interested in slips across  which 
the atomic  displacement is discontinuous 
by the lattice constant $a$. 
In this paper, by setting  $\Delta x=a$, 
we try to realize such singular objects  numerically.  
For this purpose it is convenient 
to  define  the strains and tensors on the middle points 
$(n+1/2,m+1/2)$, while the vectors@
are defined at the lattice points  $(n,m)$. 
For  a vector 
component $A$ (say, $ A=u_x$),   
$\nabla_x A$ and $\nabla_y A$ at $(n+1/2,m+1/2)$ are  defined as 
$[A(n+1,m+1)-A(n,m)+A(n+1,m)-A(n,m+1)]/2$ and 
$[A(n+1,m+1)-A(n,m)-A(n+1,m)+A(n,m+1)]/2$, respectively,   
using  $A$ at the four points 
$(n+1/2\pm 1/2,m+1/2\pm 1/2)$.  
In the same manner, we may construct 
the vector $\nabla\cdot\asigma$ at $(n,m)$ 
using the stress components at the four points 
$(n\pm 1/2,m\pm 1/2)$.  With this space discretization,  
slips consisting of a straight line segment 
become   well-defined 
if their angle with respect to 
the $x$ or $y$ axis  is  $0$ or  $\pi/4$. For 
other slip orientations,  zigzag points 
 appear along  the slip line segment 
and an extra elastic energy 
becomes  needed. 
On the other hand, if we define all the quantities on 
 $(n,m)$, slip discontinuity 
takes place over a few lattice sizes, 
but macroscopic features 
such as the stress-strain relation 
remain almost unchanged. 
Furthermore, 
we will suppose  simple shear or uniaxial deformation 
and, as will be shown in the next section, 
 the preferred slip orientation angle is 
$0$ or $\pi/4$ with respect to the $x$ or $y$ axis. 
By this reason our simple numerical scheme 
seems to be allowable at least in this first attempt.

\section{Slips} 
\setcounter{equation}{0}

In our model system fundamental flow units 
in plastic deformation  are slips composed of 
a pair of edge dislocations with opposite 
Burger vectors (dislocation dipoles). 
They are analogous to quantum vortex rings 
in superfluid helium \cite{Onukibook}. 
Their elastic structure 
far from the dislocation cores may  well be described by 
the linear elasticity theory, but nonlinear elasticity  theory 
is needed  (i)   to suppress 
 the divergence of the stress  
at  the cores and (ii)  
to stabilize  the  slips themselves 
when they  adjust to the 
crystal structure. 
Slips are not in a stationary state  
in the linear elasticity theory in the absence of 
impurities etc. which can trap dislocations,  
as will be evident in 
(3.20).  In our nonlinear  
theory,  those along the $x$ axis ($\theta=0)$
 can  be in a stationary 
metastable state if  their length 
is a multiple of  the lattice constant.  
This is  consistent with the Peierls-Nabarro 
theory  \cite{P,N}, 
which takes into account the discreteness 
of the crystal structure and gives 
a periodic Peierls potential energy   
for the position of the 
dislocation center.

\subsection{Slips in linear elasticity theory}

To begin with, let us write 
the  solution of an edge dislocation 
as ${\bi u} = b{\bi u}^{\rm Le} = (b u_x^{\rm Le} ,  b u_y^{\rm Le} )$, 
whose  Burgers vector  is 
assumed to be along the 
$x$ direction  and is written as 
${\bi b}= (b,0)$. The 
linear elasticity theory  \cite{Landau} gives   
\bea 
u_x^{\rm Le} &=& 
\frac{1}{2\pi} \bigg [ \tan^{-1} \bigg (\frac{y}{x} \bigg ) 
+\frac{1}{2(1-\nu)} \frac{xy}{x^2+y^2} \bigg ]\nonumber\\
u_y^{\rm Le} &=& \frac{-1}{4\pi(1-\nu)} 
\bigg [ (1-2\nu)  \ln \sqrt{x^2+y^2}
+ \frac{x^2}{x^2+y^2}  \bigg ]. 
\label{eq:3.1}
\ena 
In the  3D theory  the dislocation line is along the $z$ 
axis and $\nu$ is  3D Poisson's  ratio. In our 2D theory $\nu$ 
 is related to $\lambda$ in (2.23) as 
\be 
\nu = \frac{1}{2}- \frac{1}{2\lambda}. 
\label{eq:3.2}
\en 
In our simulations  
 $\lambda=1$ and hence  $\nu= 0$.  
In the linear theory  
a slip   is a  superposition of two edge dislocations 
with opposite Burgers vectors expressed as   
\be 
{\bi u}^{{\rm Ls} \pm}
 = \pm   \bigg [ {\bi u}^{\rm Le} (x-\frac{\ell}{2},y) 
-{\bi u}^{\rm Le} (x+\frac{\ell}{2},y) \bigg ] , 
\label{eq:3.3}
\en 
where $\ell$ is  the slip length 
and $b=\pm 1$ (both  in units of $a$). The slip line segment 
is between the two points $(-\ell/2,0)$ and $(\ell/2,0)$. 
The $+$  sign corresponds to 
a clockwise slip (type C), and the $-$ sign to 
a counterclockwise slip (type CC). The displacement vector 
around a slip is clockwise (counterclockwise) 
for type C (type CC) (as will be evident in Fig.5 below).
Across the line segment 
$|x|<\ell/2$ and $y=0$, the displacement is  discontinuous as 
\be 
u_x^{{\rm Ls} \pm}(x,y+0)-u_x^{{\rm Ls} \pm}(x,y-0)= \pm  1.
\label{eq:3.4}
\en    
The corresponding  strains are  written as 
\bea 
e_1^{{\rm Ls} \pm}&=& 
\frac{\pm (2\nu-1)}{2\pi(1-\nu)} \bigg 
[ \frac{y}{x_+^2+y^2} 
- \frac{y}{x_-^2+y^2} \bigg ], \nonumber\\ 
e_2^{{\rm Ls} \pm}&=& 
\frac{\mp 1}{\pi(1-\nu)} \bigg [ \frac{x_+^2y}{(x_+^2+y^2)^2} 
- \frac{x_-^2y}{(x_-^2+y^2)^2} \bigg ] ,\nonumber\\ 
e_3^{{\rm Ls} \pm}&=& 
\frac{\pm 1}{2\pi(1-\nu)} \bigg [ \frac{x_+(x_+^2-y^2)}{(x_+^2+y^2)^2} 
- \frac{x_-(x_-^2-y^2)}{(x_-^2+y^2)^2} \bigg ] \nonumber\\ 
 && \pm \delta(y)\Theta(\ell^2/4-x^2), 
\label{eq:3.5}
\ena
where $x_{\pm} = x \mp  \ell/2$, and 
  $\Theta(\zeta)$ is the step function being equal to 1 for $\zeta>0$ 
and to $0$ for $\zeta<0$.  There appears no $\delta$-function 
in the dilation and tetragonal strains. 
The strains diverge at the cores where the linear elasticity 
theory breaks down.

\subsection{Slips in nonlinear elasticity theory}

Next we numerically construct the corresponding slip solution 
in our nonlinear elasticity theory for integer  $\ell$. 
That is, by  starting  with 
the linear solution ${\bi u}^{{\rm Ls} \pm}$ in 
(3.3) at $t=0$ and neglecting the 
random stress ${\asigma}_R$, 
we integrate (2.16) and (2.17) to 
seek the steady solution 
${\bi u}^{{\rm s}\pm}$  attained after 
transient relaxation.  That is, 
\be 
{\bi u}^{{\rm s}\pm}(x,y)=\lim_{t\rightarrow\infty} {\bi u}(x,y,t) ,
\label{eq:3.6}
\en 
where ${\bi u}(x,y,0) = {\bi u}^{{\rm Ls} \pm}(x,y)$. 
The core points of the linear solution,  
where the strains diverge,  are placed at 
middle points $(n+1/2,m+1/2)$  of  the 
mesh  of integration at $t=0$. 
The limit  
${\bi u}^{{\rm s}\pm}$ is a steady metastable 
solution satisfying the mechanical 
equilibrium condition,
\be 
\nabla\cdot\asigma={\bi 0}.
\label{eq:3.7}\en 
It nearly coincides with the linear solution 
${\bi u}^{{\rm Ls} \pm}$ far from the 
dislocation  cores ($|x\pm \ell/2|^2 +y^2 \gs 3$ in our case)    
and keeps  to satisfy the slip condition 
(3.4).    The stain and stress  components 
 calculated from ${\bi u}^{{\rm s}\pm}$ 
are  finite even in the core regions. 
In fact, in the presence of  a slip 
along the $x$ direction,  the maximum values attained by  
$|e_1|$, $|e_2|$, $|e_3|$, 
and $|\sigma_{xy}|$ in the core regions  
are about $0.18$, $0.08$, $1.1$, 
and $0.1$, respectively. 
In Fig.4 we show  the $x$ component $u_x= u_x^{{\rm s}+}$ 
of  a  clockwise slip with $\ell=20$ in the unstrained condition, 
which is discontinuous by 1 across the slip segment.

The elastic energy of a slip in our nonlinear theory 
is then of central  importance. We will neglect the 
Peierls potential energy for the time being. 
We generally assume that a slip line is oblique to the $x$ axis 
making an angle of $\varphi$ and is under externally applied 
strains,
\be  
\av{e_3} =\gamma,\quad \av{e_2} =\epsilon.
\label{eq:3.8}
\en 
We assume $\av{e_1} =0$.  The average displacement is given by 
$\av{u_x}= \gamma y + \epsilon x/2$ 
and $\av{u_y}= -\epsilon y/2$. 
 Here $\av{\cdots}$ is the space average. 
The slip energy to create a single slip 
is defined by 
\be 
 F_{\rm slip} = F -F_0 =\int d{\bi r} (f_{\rm el} - f_{\rm el} ^0), 
\label{eq:3.9}
\en 
where $f_{\rm el} $  is the elastic energy density in our nonlinear theory 
with one slip   calculated numerically,    
and  $f_{\rm el} ^0= \Phi(\gamma,\epsilon)$ is that in  
 a homogeneously  strained state.  
In the unstrained condition ($\gamma=\epsilon=0)$,   the expression  
$F_{\rm slip}  = \ln \ell/{2\pi(1-\nu)}$  (in units of $\mu_0a^2$) 
is well known \cite{energy}.

Dislocation  motions  perpendicular to the 
slip line  ({\it climb}  motions) 
create a large number of defects 
 ($\propto \ell$) and the energy needed is very large, so they 
  may be neglected \cite{climb}.  
On the other hand, dislocation motions along the slip line ({\it glide} motions) 
involve displacements  of a relatively small number of particles 
(of order 10 for a unit-length motion 
as can be seen in the inset of 
Fig.9) and hence play a major role  in plastic flow.
A force  $\bi{\cal F}=({\cal F}_x,{\cal F}_y)$  
acting on a dislocation under an applied 
stress $\{ \sigma_{ij}^{\rm ex} \}$ 
may be calculated using the 
Peach-Koehler theory \cite{Friedel,Nabarro,Peach}. 
 For ${\bi b}= (b,0)$  along the $x$ axis, 
the components of the force are written as 
\be 
{\cal F}_x= -\sigma_{xy}^{\rm ex}  b , \quad 
{\cal F}_y= \sigma_{xx}^{\rm ex}  b . 
\label{eq:3.10}
\en 
(i) For a slip  along the $x$ axis, 
${\cal F}_y$ is canceled 
by the force due to an additional elastic 
deformation in the surrounding medium. 
If the dislocation on the left is fixed 
and that on the right is moved by $\delta\ell$ 
along the $x$ axis, the change of 
$ F_{\rm slip} $ is equal to ${\cal F}_x\delta\ell$.  
Here the Peierls force is neglected. Therefore,  
\be 
\frac{\p}{\p \ell} F_{\rm el}  = {\cal F}_x= -\sigma_{xy}^{\rm ex} b.
\label{eq:3.11}
\en 
The  $\sigma_{xy}^{\rm ex} $ consists of 
the stress produced by the dislocation on the left 
and the externally applied stress. 
For small $\gamma$ we obtain 
\be 
 F_{\rm slip} = \frac{\ln \ell}{2\pi(1-\nu)}  
 \mp   \gamma \ell  , 
\label{eq:3.12}
\en 
where $-$ is for  type C and $+$ is for type CC. 
(ii) For general angle $\varphi$ of the slip 
with respect to the $x$ axis,  
we rotate the reference frame by $\varphi$ 
to obtain 
the shear strain 
 $\gamma'= \gamma \cos 2\varphi- \epsilon \sin 2\varphi$
in the new reference frame from (2.2). 
Therefore, 
\be 
 F_{\rm slip} = \frac{\ln \ell}{2\pi(1-\nu)}  
\mp   (\gamma \cos 2\varphi- \epsilon \sin 2\varphi) \ell  . 
\label{eq:3.13}
\en

If $\varphi$ is varied in (3.13), 
 $F_{\rm slip}$ is minimized for 
\be 
\varphi= \frac{1}{2}(n \pi - \alpha) \quad (n=0,1,2,\cdots), 
\label{eq:3.14}
\en 
where $n$ is even for type C 
and odd for type CC, and 
$\alpha$ is determined by 
\be 
\cos\alpha= \frac{\gamma}{\sqrt{\gamma^2+\epsilon^2}}, \quad 
\sin\alpha= \frac{\epsilon}{\sqrt{\gamma^2+\epsilon^2}}. 
\label{eq:3.15}
\en 
For  simple shear deformation with $\gamma>0$ and $\epsilon=0$,  
 the most favorable slip orientation with the lowest 
$F_{\rm slip} $ is 
$\varphi=0$ for type C and $\varphi=\pi/2$ for type CC.
For  uniaxial stretching  with $\gamma=0$ and $\epsilon>0$,  
it is given by 
$\varphi=-\pi/4$ for type C and $\varphi=\pi/4$ for type CC, 
in agreement with the experiments 
\cite{Spaepen,Argon,Spaepen_review,Kimura,Crist,granular,granular1}.

As will be shown in the appendix,   
the slip directions 
determined by  (3.14) are perpendicular to 
the wave vectors which minimize 
the  angle-dependent 
sound velocity following from 
our linearized dynamic equations,  
or perpendicular to the {\it softest}
 directions for  the sound modes. 
This coincidence is obtained 
with the aid of the isotropic behavior (2.8), so it is not 
a general result for anisotropic solids. We remark that 
previous theories of strain localization 
\cite{granular1,Rice,Langer1} 
are  based on 1D analysis, 
where all the quantities vary only  
in one direction $\bi n$ normal to the plane of the band, 
and  reduce to linear stability analysis 
for small amplitude perturbations for  the  
wave vector  in the direction of $\bi n$.

In Fig.5 we display the displacement 
vector  $\bi u$  around   type C  and  type CC slips calculated in the 
unstrained condition. The slips are oriented  in 
  the most favorable directions in shear deformation in (a) 
and in uniaxial stretching in (b).  Away from the slips the directions of 
$\bi u$  continuously change to those of  the  
 macroscopic deformation  
supposed to be applied. In Fig.6 the slip energy $F_{\rm slip} $ 
of type C slips with  $\ell=10, 20,$ and 30 is shown as a function 
of the applied shear strain $\gamma$.  For $|\gamma| \ls 0.05$ 
we confirm (3.12) with the $-$ sign.  For larger  $\gamma$  
the linear relation $\p F_{\rm slip} /\p \gamma$ =const.
 does not hold.   For $\gamma<\gamma_{{\rm c}1} (\sim -0.1)$  
or for  $\gamma>\gamma_{{\rm c}2} (\sim 0.1)$,  
a steady  metastable  solution becomes nonexistent 
and, as a result,  the slip grows up to the system length 
or shrinks to vanish in the simulation.  
In Fig.6 the slip energy  
$f_{\rm el} -f_{\rm el} ^0$ with $f_{\rm el} ^0= \Phi(\gamma,0)$ 
is displayed for the three values $\gamma=0, 0.065$, and $-0.04$.   
Interestingly,  the elastic energy density in the 
middle region between the two dislocations at the ends   
is decreased for positive $\gamma$ and 
increased for negative $\gamma$, giving rise to 
the  contribution $-\ell \gamma$ in the slip energy. 
In  the core region,  $f_{\rm el} $ has two peaks 
and is rather insensitive to applied strains in our model.

Furthermore, in Fig.8 we numerically demonstrate  coincidence of 
$\p F_{\rm slip} /\p \gamma$ and the space integral of 
$\sigma_{xy}-\gamma$ for a single isolated slip in 
 simple shear deformation. 
This relation may be obtained from 
(2.1). If $\gamma$ is increased by an infinitesimal amount $\delta\gamma$, 
the  change 
of $F_{\rm slip} $ is written in terms of the incremental displacement 
$\delta { u}_i$ as 
\be 
\delta F_{\rm slip} = \int d{\bi r} \sum_{ij} [\sigma_{ij} -\sigma_{ij}^0]
\frac{\p}{\p x_j} \delta u_i ,
\label{eq:3.16}
\en 
where $\{\sigma_{ij}^0\}$ is the stress in the homogeneous 
state  and  use is made of 
the relation for the space average 
$\av{{\p} \delta u_i/{\p x_j}}=  
\delta_{ix}\delta_{jy}\delta\gamma$. 
 The deviation  $\delta { u}_i$ 
consists of the applied displacement 
change $ \delta_{ix}y\delta\gamma$  
and the induced deviation $\delta {u}'_i$  
localized near the slip. However, the contribution from 
$\delta u'_i$ vanishes in (3.16) 
from the mechanical equilibrium 
condition (3.7).
Thus, 
\be 
\frac{\p}{\p \gamma}F_{\rm slip} (\ell,\gamma)
= \int d{\bi r} [\sigma_{xy} -\sigma_{xy}^0]. 
\label{eq:3.17}
\en    
In Fig.8 $\gamma$ is in the range 
$|\gamma | \ls 0.1$, so 
 $\sigma_{xy}^0 \cong \gamma$ holds excellently   
for our elastic energy.  
Similarly, the counterpart of (3.17) 
in the case of uniaxial stretching is written as 
\be 
\frac{\p}{\p \epsilon}F_{\rm slip} (\ell,\epsilon)
= \frac{1}{2} 
\int d{\bi r} [\sigma_{xx} -\sigma_{yy} - \sigma_{xx}^0+\sigma_{yy}^0]. 
\label{eq:3.18}
\en    
The  relations (3.17) and (3.18) 
 hold for any strain amplitudes as long as a steady slip 
solution can be obtained from  (3.6), while (3.12) or (3.13) 
is valid only for very small strains ($\ls 0.05$ 
in our case).

\subsection{Peierls potential energy}

We continue to consider a slip 
along the $x$ axis under simple shear deformation, but 
the slip length $\ell$ here can be noninteger. 
For general $\ell$ we modify (3.12)  as  
\be 
 F_{\rm slip} = \frac{\ln \ell}{2\pi(1-\nu)}  
 \mp   \gamma \ell  + U_{\rm PN} (\ell), 
\label{eq:3.19}
\en 
where  $U_{\rm PN} (\ell)$ represents 
the  Peierls potential energy being zero 
for integer $\ell$ and positive  
for noninteger $\ell$.  
The force acting on  the slip along the glide direction 
is then given by 
\be 
 \frac{\p}{\p \ell}
F_{\rm slip} = [{2\pi(1-\nu)}]^{-1} \frac{1}{\ell}   
 \mp   \gamma + \frac{\p}{\p \ell} U_{\rm PN} (\ell).  
\label{eq:3.20}
\en 
This force  vanishes  for a stationary slip. 
In the realm of linear elasticity 
 theory, where the crystal structure is smoothed out, 
 the  Peierls potential energy is nonexistent 
and we are led to the following conclusion: 
A  type CC slip  shrinks 
for any $\ell (>1)$,  
but  a type C slip  shrinks 
for $\ell<\ell_c^{\rm L}$ and  expands for $\ell>\ell_c^{\rm L}$. 
Here we assume $\gamma>0$ and the 
critical length $\ell_c^{L}$ in the 
linear elasticity theory is determined  by 
\be 
\ell_{\rm c}^{\rm L}=  1/[{2\pi(1-\nu)}\gamma].
\label{eq:3.21}
\en 
If $\ell$ is fixed, we obtain 
a critical shear strain $\gamma_{\rm c}^{\rm L} 
= 1/[{2\pi(1-\nu)}\ell]$.  In our 
nonlinear theory, however, this 
critical length (or strain) loses its 
physical  relevance.

In our simulations 
(without the random stress) 
slips can be stationary suggesting the 
existence of   the  Peierls potential energy.  
To show this,  we numerically create   two 
type C slips along the $x$ axis obtained in the limit  (3.6); one is  
in the range  $0 \le x \le  20$ with length 20, and 
the other is in the range 
 $-1 \le x \le 20$ with length 21. Here the positions  
of the dislocation core on the left are different by $1$ 
but  those on the right coincide. 
As a result, the corresponding 
displacements ${\bi u}_{20}$ and ${\bi u}_{21}$ 
are different only near the core region 
on the left, as can be 
seen in  the inset of Fig.9. Note that  the difference 
${\bi u}'= {\bi u}_{21}- {\bi u}_{20}$  is  the displacement 
realized  when  the shorter slip  grows  
into the longer one.  
Now, we calculate  $F$    for the 
interpolated displacement, 
\be 
{\bi u}_\ell=  (1-\alpha){\bi u}_{20}+ 
\alpha {\bi u}_{21} ={\bi u}_{20}+ \alpha{\bi u}'.
\label{eq:3.22}
\en  
As the slip energy at 
$\ell=20+\alpha$ ($0<\alpha<1$), 
 $F_{\rm slip}(\ell)=F-F_0$ 
is determined as in  (3.9).    
The  $F_{\rm slip} (\ell)$ here 
 depends on the displacement path 
connecting ${\bf u}_{20}$ 
and ${\bf u}_{21}$.

In Fig.9   the resultant energy difference 
 $\Delta F_{\rm slip}=F_{\rm slip}(\ell)-F_{\rm slip}(20)$ is shown. 
 From this calculation the Peierls potential 
cannot be determined  uniquely but, if we  define  
\be 
U_{\rm PN} (\ell)=  F_{\rm slip}(\ell)- 
(1-\alpha)F_{\rm slip}(20) - \alpha F_{\rm slip}(21) , 
\label{eq:3.23}
\en 
we  can know its behavior in the range 
$0<\alpha=\ell-20<1$ directly from  Fig.9. 
We recognize that $F_{\rm slip}(\ell)$ 
takes  a maximum at  $\ell=\ell_{\rm max}$ 
between $20$ and $21$  and 
takes   local minima at $\ell=20$ and 21. 
It follows that the stable force-balance 
condition  $ \p F_{\rm slip} /\ell=0$  with $ \p^2 F_{\rm slip} /\ell^2>0$ 
holds at $\ell=20$ and 21.
The maximum of   $U_{\rm PN}$ 
is about  0.03 for $\gamma=0$. Fig.9 also indicates that 
$\ell_{\rm max}  \rightarrow 21$  as  
$\gamma \rightarrow \gamma_{{\rm c}1}  (\cong -0.09)$ 
and 
$\ell_{\rm max}  \rightarrow 20$ as 
$\gamma \rightarrow \gamma_{{\rm c}2}  (\cong  0.12)$.
For  integer $\ell$ and noninteger $\ell'$,  
 the critical strains,  
$\gamma_{{\rm c}1} (\ell)$ and $\gamma_{{\rm c}2} (\ell)$, 
with respect to  shrinkage 
and expansion  satisfy the following:  
(i) $\p F_{\rm slip} (\ell')/\p \ell' >0$ 
for any $\ell'$ smaller than $\ell$ 
for $\gamma<\gamma_{{\rm c}1}(\ell) $, 
(ii) $\p F_{\rm slip} (\ell')/\p \ell' <0$ for any $\ell'$ larger  than $\ell$ 
for $\gamma>\gamma_{{\rm c}2}(\ell) $, 
and (iii) $F_{\rm slip} (\ell)$ is locally minimum 
at integer length $\ell$ for  $\gamma_{{\rm c}1}(\ell)
 <\gamma<\gamma_{{\rm c}2}(\ell) $.  
These are consistent 
with the positions of 
 the instability points in Fig.6.

\section{Plastic flow} 
\setcounter{equation}{0}

In this section 
we induce  deformation at 
a constant strain rate, $\dot{\gamma}$ or $\dot{\epsilon}$, 
or cyclic shear deformation 
for  $t\ge 0$ in the presence of the random stress 
tensor ${\asigma}_{\rm R}$  with 
$\epsilon_{\rm th} =0.1$ 
(except for the curve (b) in Fig.22 
where $\epsilon_{\rm th} =0.25$). 
At $t=0$, the values of $\bi v$  at the lattice points  
are Gaussian random numbers 
with variance $0.01$. 
The shear stress 
and   the normal stress difference 
in the following figures are the space averages,  
$\av{\sigma_{xy}}$ and    $\av{\sigma_{xx}-\sigma_{yy}}$, 
respectively.

\subsection{Shear deformation}

In Fig.10 we  show the stress-strain curves 
obtained by  integration of  (2.16) and (2.17) 
under  a constant shear rate $\gdot$. 
At small $\gamma(t)=\gdot t$ these  curves 
first follow the curve in  
the homogeneous case, 
\be 
\sigma_{xy} \cong A_3(\gamma) = 
  \frac{1}{\sqrt{3}\pi} \sin (\sqrt{3} \pi \gamma) , 
\label{eq:4.1}
\en 
where $A_3$ was introduced in  (2.15).
For the  curve of $\gdot=10^{-4}$ and that of  $\gdot=10^{-3}$ 
(with the higher peak)  
we set   ${\bi u}={\bi 0}$ at $t=0$ 
(supposing  a perfect crystal).  
They approach the elastic instability point 
$\gamma = \sqrt{3}/6\cong 0.289$,  where 
$\p \sigma_{xy}/\p \gamma$ tends to vanish 
and  softening with respect to further 
shear deformation occurs.  
Then the shear stress  drops sharply 
after the peak with 
catastrophic formation of slips. 
See the two snapshots of 
$\delta e_3=e_3-\gamma$  in the figure. 
The orientations of the slips 
are the most favorable ones 
with $\varphi=0$ and $\pi/2$ as determined in  (3.14). 
For $\gdot=10^{-4}$ the slip formation is triggered  
at a smaller strain ($\cong 0.25$) 
and the spacing 
between the  slips is a few times wider than 
in the case of $\gdot=10^{-3}$. In our simulation 
 the slip spacing 
depends on the shear rate $\gdot$ in the plastic 
flow regime with the other dimensionless 
parameters held  fixed. 
For the other curve of  $\gdot=10^{-3}$ 
(with the lower peak)  
we put four slips with length 20 at $t=0$,  
as will be illustrated in Fig.13 below in  detail.  
This curve indicates 
that the overshoot in the 
stress-strain relation  is   weakened 
 by the  initially preexisting defects. 
In Fig.11 
we display mechanically unstable points (dots)
for   $\gamma=0.29$, $0.30$, and $0.40$ 
at $\gdot=10^{-3}$, where 
at least one of  the eigenvalues $\lambda_1$ and $\lambda_2$ 
of the matrix $\{\Phi_{\alpha\beta}\}$ defined below (2.7) 
is negative. We recognize that 
the stability condition  $(\lambda_1>0$ and $\lambda_2>0$) 
are satisfied at most points in plastic flow. 
Fig.12 displays a snapshot of the displacement vector $\bi u$ 
in a $1/4$ region of the total system at $\gamma=0.4$ 
with $\gdot=10^{-4}$.  We can see a number of slips 
and  bands (aggregates of slips here), 
where  $\varphi=0$ for type C 
and   $\varphi=\pi/2$ for type CC. 
 The large horizontal 
shear band in the lower part is particularly conspicuous, 
where the band  thickness and the 
discontinuity of $u_x$ across it    
are both increased up to about 3-4.     
We recognize that  elementary slips (dipoles of edge dislocations) 
tend to be created around preexisting ones, 
yielding  thicker  shear bands. 
Similar thick shear bands have been  
observed in previous simulations 
\cite{Falk,Bulatov}.

In Fig.13 we follow time-development of $\delta e_3$ 
and $\delta f_{\rm el} = f_{\rm el} -\av{f_{\rm el} }$ at $\gdot=10^{-3}$
in the presence of four slips at $t=0$, where 
the favorable slips (in black in the upper 
figure at $\gamma=0$)  grow and 
the unfavorable ones (in white)  shrink as $\gamma$ increases. 
In the lower snapshots the elastic energy density 
deviation $\delta f_{\rm el} =f_{\rm el} - 
\av{f_{\rm el} }$ is shown, where 
the black dots represent the dislocation cores. 
At $\gamma=0.176$ we can see that 
the energy density between the two 
dislocations at the slip ends is decreased  for 
 the favorable slips and is increased 
for the unfavorable slips, in accord with (3.13) and Fig.7.
 At the plastic flow regime 
$\gamma=0.387$ the initial  favorable slips 
grow into thick layers  where the  
dislocation density is  very high.

Next we apply a cyclic shear deformation,  
where $\gdot(t)=10^{-3}$ in the time regions 
 $n t_{\rm p}<t<(n+1/2)t_{\rm p}$ 
and $\gdot(t)=-10^{-3}$ in the time regions $(n+1/2)
t_{\rm  p}<t<(n+1)t_{\rm p}$. 
We choose  $t_{\rm p}=1000$, so the maximum 
of $\gamma(t)$ is 0.5 and the minimum is 0. 
For the first two cycles $(n=0$ and 1), 
Fig.15 shows the stress-strain curve,
 while Fig.16 shows the average elastic 
energy density $\av{f_{\rm el}}$ 
with a snapshot of $f_{\rm el} $ at the point A where 
the average stress vanishes. 
As salient features,  we notice 
(i) residual strain at  vanishing 
stress, (ii) the shear stress becomes 
negative at the end of the first 
cycle, (iii) no   overshoot
 in the stress and the elastic energy 
 from  the second cycle, 
and (iv) that the stress and the elastic energy 
take roughly constant values characteristic of 
well-developed plastic flow 
in the region $0.25 \ls \gamma (t) <0.5$ for increasing $\gamma(t)$  
and in the region $0 < \gamma (t) \ls 0.25$ for decreasing $\gamma(t)$.  
Thus we can see  significant hysteresis behavior.  
In MD simulations of low $T$ 
glasses, similar  stress-strain curves 
under step-wise  strain rates  
have  been obtained (but without overshoot  behavior because of 
disordered initial states)  
\cite{Deng,Langer}.

We also 
notice that at the points A,B, and C, where the average 
stress vanishes as in Fig.14, 
the curves of $\av{f_{\rm el} }$ in Fig.15 
are locally minimum. 
This suggests that 
 the strain $\gamma(t)$ consists of 
an elastic strain $\gamma_{\rm el}$  
and an slip strain 
\be 
\gamma_{\rm s}= \gamma- 
\gamma_{\rm el}.
\label{eq:4.2}
\en 
Roughly speaking, the elastic strain outside the dislocation 
cores should give to the average stress, 
while the slip strain 
is caused  by the jumps of $u_x$ across the slips.
To be more quantitative, we  define $\gamma_{\rm el}$ as 
\be 
A_3(\gamma_{\rm el})  = \av{\sigma_{xy}}. 
\label{eq:4.3}
\en  
The elastic energy density stored is then the  
sum of the elastic energy density 
and  the defect energy density. 
We define the  average 
defect energy density by 
\be 
f_{\rm D}= \av{f_{\rm el}}- \Phi(\gamma_{\rm el},0),
\label{eq:4.4}
\en  
where $\Phi$ is defined by (2.5). 
For small $\gamma_{\rm el}$   
we have $\gamma_{\rm el}  \cong 
 \av{\sigma_{xy}}$ and  
$\Phi (\gamma_{\rm el},0) 
\cong \gamma_{\rm el}^2/2$. 
Hence, in the vicinity of  the points 
where $\av{\sigma_{xy}} =0$,
$\av{f_{\rm el}}(t)$ should take a minimum,  
 provided that  $f_{\rm D}(t)$ changes slowly there.  
To confirm these arguments, we plot $f_{\rm D}(t)$ 
in Fig.16. For $\gamma (t) \ls 0.3$ 
in the first cycle, however, 
it represents 
the elastic energy due to the inhomogeneous 
fluctuations of the local strains (mainly due to 
$\delta e_3$) with the peak height at 
0.012 (not shown in the figure). 
After this initial period,  
$f_{\rm D}(t)$ is 
 in a range of $0.004-0.005$ 
reasonably representing 
 the elastic energy due to the defects 
produced in plastic flow. 
The energy variance 
$\sqrt{\av{(\delta f_{\rm el})^2}}$ 
is also about $0.005$ in plastic flow 
obviously due to the discrete 
nature of dislocation cores. 
For $\gdot=10^{-4}$, on the other hand,   
$f_{\rm D}(t)$ is almost  constant 
around $0.002$ in plastic flow.

\subsection{Uniaxial stretching}

In Fig.17 we show the normal stress vs  strain at 
 constant  strain 
rate $\dot{\epsilon}$ for $t\ge 0$. 
The characteristic features are very similar to those in Fig.10. 
At small $\epsilon(t)=\dot{\epsilon} t$ these  curves 
first follow the curve in  
the homogeneous case, 
\be 
\sigma_{xx}-\sigma_{yy}\cong 2A_2(\epsilon)= 
\frac{2}{{3}\pi}  [ \sin (\pi \epsilon) +\sin (2\pi \epsilon)  ], 
\label{eq:4.5}
\en 
where $A_2$ was introduced in  (2.15). 
For the  curve of $\dot{\epsilon}=10^{-4}$ and that of  
$\dot{\epsilon}=10^{-3}$ 
(with the higher peak)  
we set   ${\bi u}={\bi 0}$ at $t=0$.  
They approach the elastic instability point 
$\epsilon = \pi^{-1} \cos^{-1}[(\sqrt{33}-1)/8]\cong 0.298$. 
After the catastrophic formation of slips, 
the slip orientation angles are $\pm \pi/4$ 
with respect to the stretched direction  
 in agreement with the  experiments 
\cite{Spaepen,Argon,Spaepen_review,Kimura,Crist,granular,granular1}   
and the simulations \cite{Falk,Bulatov}. 
The discontinuity across  the  slip lines  appears   in the 
tetragonal strain $e_2$,  which 
can be seen  in the two snapshots of $\delta e_2= e_2-\epsilon$.
In Fig.18  the displacement vector $\bi u$ 
in a $1/4$ region of the total system at $\epsilon=0.46$ 
with $\dot{\epsilon} = 10^{-3}$ is displayed, 
where the orientations of the slips 
are the most favorable ones 
with $\varphi=\pm \pi/4$  as determined below (3.14).  
In Fig.19  four slips are placed at $t=0$, exactly 
as in Fig.13,  and time-development of  
$e_3$, $\delta e_2$, and $\delta f_{\rm el} $ are followed.
  Here, however, these initial slip orientations 
are  marginal in the sense that $F_{\rm slip} $ 
in (3.13) is independent of $\epsilon$. We find that 
a dislocation pair  is  newly created near each  
end of the preexisting slips to glide 
  in the favorable 
directions with $\varphi=\pm \pi/4$. In the plastic flow regime 
at $\epsilon =0.4$,  they grow into thick layers 
containing high-density dislocations.

Next we apply  cyclic uniaxial stretching,  
where $\dot{\epsilon}(t)=10^{-3}$ in the time regions 
 $n t_{\rm p}<t<(n+1/2)t_{\rm p}$ 
and $\dot{\epsilon}(t)=-10^{-3}$ in the time regions $(n+1/2)
t_{\rm  p}<t<(n+1)t_{\rm p}$. 
We choose  $t_{\rm p}=1000$, so the maximum 
of $\epsilon(t)$ is 0.5 and the minimum is 0. 
Fig.20 shows the stress-strain curve 
for the first four cycles, 
where it is nearly periodic from the second cycle. 
As in the shear deformation case 
we introduce the average elastic 
strain $\epsilon_{\rm el}$ by 
\be 
2A_2(\epsilon_{\rm el}) 
=\av{\sigma_{xx}-\sigma_{yy}}
\label{eq:4.6}
\en 
and  the average defect  
energy density ${f_{\rm D}}$   by 
\be 
f_{\rm D}= \av{f_{\rm el}}- \Phi(0,\epsilon_{\rm el}). 
\label{eq:4.7}
\en 
where $\Phi$ is defined by (2.5). The average slip strain 
is given by $\epsilon_{\rm s}= \epsilon-\epsilon_{\rm el}$. 
For small $\epsilon_{\rm el}$   
we have $2\epsilon_{\rm el}  \cong 
 \av{\sigma_{xx}-\sigma_{yy}}$ and  
$\Phi (0,\epsilon_{\rm el}) 
\cong \epsilon_{\rm el}^2/2$. 
Fig.21 shows  the average elastic 
energy density $\av{f_{\rm el}}$ 
and the defect energy density 
$f_{\rm D}$. The latter is 
in a range of $0.008-0.009$ in plastic flow. 
The  inset of Fig.21
 displays  a snapshot of $f_{\rm el} $ at the point A 
in the first cycle, where the average normal stress difference  vanishes.  
Comparing it with the snapshot of  $f_{\rm el} $ in 
Fig.15 under shear deformation, we notice 
 a considerable difference 
in the spatial anisotropy of 
the dislocation distribution between the two cases.

\subsection{Strain-induced disordered states}

In the plastic flow regime  
dislocations  are proliferated   
and a structurally  disordered state is 
realized.  This effect may be called {\it strain-induced 
disordering}. We mention a simulation 
by  Ikeda {\it et al.}  \cite{Ikeda}, who 
 applied a tensile strain 
to a 3D model to induce  
a change from a perfect crystal to 
an amorphous solid. 
 In Fig.4 in our previous work \cite{OnukiJ} 
we   switched off a shear flow  
(a) before the peak time of the stress, 
(b)  just after the peak time, and 
(c) in  well-developed plastic flow.   Affine 
 deformation in (a) was  maintained, while 
no appreciable time evolution was detected  after 
transients in (b) and (c). 
This means that the structurally 
disordered  states are metastable.

In Fig.22, we show 
extremely slow relaxation of the 
average energy density $\av{f_{\rm el}}(t)$,  
where a shear flow with $\gdot=10^{-3}$ is stopped 
at time $t_{\rm A}\cong  600$ (at the point A in Fig.14) and 
the system is relaxed 
for a time period of $t_{\rm w}= 10^5$. 
Here we set $u_x= \gamma_{\rm A}L_0$ and $u_y=0$ 
 at the top $y=L_0$ 
with $\gamma_{\rm A}\cong 0.40$, together with 
${\bi u}={\bi 0}$ at the bottom. 
The dislocation distributions here closely 
resemble that  in the inset of Fig.15 
(are not identical because they are obtained from 
different runs). In the upper curve (a) the noise strength 
 $\epsilon_{\rm th}$ in (2.25) 
is $0.1$ as in the previous 
simulations in this section, 
where each maximum in the initial stage 
corresponds to an energy increase 
accompanied with a  configuration change 
around a dislocation core. 
From (a) the typical energy increase is of order 
$10^{-5}N^2 \sim 0.1$ for each event. 
However, there is no appreciable 
relaxation for $t\gs 3\times 10^3$. 
In (b) we increase the noise strength 
$\epsilon_{\rm th}$ to $0.25$ to obtain 
a larger  energy decrease  in $\av{f_{\rm el}}(t)$ 
with  a larger thermal noise superimposed. 
In these  cases  only a small number of 
 configuration changes $(\sim 10$) occur around 
dislocation cores even for  $t_{\rm w}= 10^5$, so 
the effect of the  structural relaxation 
 is negligible on the macroscopic level (no aging effect). 
In fact,  the stress-strain 
curves after  switching-on of the shear flow 
at $t=t_{\rm A}+t_{\rm w}$ 
are   almost  independent of  $t_{\rm w}$ 
if plotted as a function of $\Delta \gamma(t)= \dot{\gamma}(
t-t_{\rm A}-t_{\rm w})$. Interestingly, 
they exhibit a rounded  peak at $\Delta \gamma= 0.15$ 
 as can be seen  in the inset. In addition, 
in the waiting time region 
($t_{\rm A}<t<t_{\rm A}+t_{\rm w})$, 
the average $\av{\sigma_{xy}}(t)$  
fluctuates  in time around 0, but its 
 noise amplitude is only of order $0.002$ for (a) 
and $0.004$ for (b) and the stress is nearly fixed.

\section{Summary and concluding remarks}

In summary, we have presented 
a nonlinear elasticity theory 
taking into account of  periodicity of 
the elastic energy density with respect to 
the shear and tetragonal strains, $e_3$ and $e_2$. 
It has the symmetry of 
 the 2D triangular lattice but  is surprisingly 
isotropic in the $e_3$-$e_2$ plane.
We summarize 
 our main results together with 
some comments.

(i) We have  numerically examined 
the slip structure 
as a function of its length $\ell$, its angle $\varphi$, and  
applied  strains $\gamma$ and $\epsilon$.  
In external strain field, slips should 
appear in the orientations  minimizing  the slip 
energy $F_{\rm slip} $ and they should grow into shear bands observed 
in  previous experiments under 
external load in various materials. 
The snapshots of the displacement 
vector in  the plastic flow regime,  
Fig.12 for shear deformation and 
Fig.18  for uniaxial stretching, 
 most unambiguously  illustrate the physical processes 
taking place.  We remark that  the previous 
experiments have mostly 
been performed under uniaxial (biaxial or triaxial) 
deformation, so future 
 experiments with shear deformation 
($\av{u_x}= \gamma y$ and $\av{u_y}=0$) 
 should  be informative.   On the other hand, 
in crystalline solids  with 
strong crystal anisotropy, 
the slip planes  are  parallel to particular crystal  planes 
 \cite{crystal}. 
In our simulations 
we may  obtain slips as steady solutions 
of our dynamic model due to the presence of 
the Peierls potential energy.  
However, they cannot be stationary 
for  $\gamma<\gamma_{{\rm c}1}  \sim -0.1$ 
or  $\gamma>\gamma_{{\rm c}2}  \sim 0.1$
 in shear strain $\gamma$, where the 
potential minima  disappear 
and the slips shrink or  grow.

(ii) We have examined plastic flow 
by applying a constant strain rate at $t=0$. 
If there is no initial disorder, 
the stress-strain curve exhibits a pronounced overshoot with a peak stress 
of order  $10\%$ of  the shear modulus. 
However, in the presence of  initial disorder, 
the overshoot is weakened or even erased,  
as revealed by simulations with initial four slips in Fig.10, 
 under cyclic shear in Figs.14 and 20, and after staying at  
the  zero-stress point A  in the inset of Fig.22. 
In accordance  with these findings, previous  simulations 
performed with various  initial states have 
demonstrated  sensitive dependence of the overshoot behavior 
on the quenching conditions \cite{Bulatov,Sti}.  
In addition, a number of previous 
 simulations  have reported  either of existence 
or nonexistence of the stress peak.  
In some real glassy systems including  polymers, 
overshoot behavior has  been widely observed 
\cite{Argon_review,Simons,Chen,Crist}. 
In an amorphous metal, 
the stress increased monotonically 
to a steady-state value for  
slow strain rate ($\dot{\epsilon} < 10^{-3}$s$^{-1}$), 
while a maximum appeared 
at $\epsilon \sim 0.06$ 
for  large strain rate ($\dot{\epsilon} >5\times 10^{-3}$s$^{-1}$) 
\cite{Chen}.

(iii) As illustrated in Figs. 16 and 21 we have 
divided the total elastic energy into the 
affine part and the defect part following  the definitions 
(4.4) and (4.7). In our model, once defects are created, 
 the defect elastic energy is rather weakly 
dependent on the  deformation history. 
On the basis of this division, 
we may easily understand the characteristic 
features of   the 
stress-stain curves in the cyclic straining 
mentioned in the previous section. 
In future microscopic simulations on glassy materials, 
this kind of  energy  division 
should be informative at relatively small strains, where  
we are  interested in 
the defect conribution  
to   measurable quantities such as 
the specific heats.  
Here we mention microscopic calculations of  the average 
potential energy 
per particle $\av{e}$ in supercooled states  under 
 shear  \cite{Sti,Lacks,Tar}, where  $\av{e}$ was  
increased considerably by shear  flow above the initial value in 
 quiescent states. 
It is remarkable that, while  
  $\av{e}$ in quiescent states sensitively depends on 
the quenching history (aging effect) \cite{Sti,Kob},  
it becomes uniquely determined  in shear flow for  
shear rates larger than the inverse structural relaxation time 
$\tau_\alpha^{-1}$. This is  because sheared systems 
are effectively  driven away from the glass transition  \cite{Yamamoto}. 
 This effect would be consistent with 
our result that $f_{\rm D}(t)$ is kept  nearly 
constant in plastic flow as in Figs.16 and 21.

(iv) In our simulations slips emerge as  
long straight lines,  as shown in the snapshots of $e_3$ or $e_2$.  
In our model  
the crystal order is not broken over 
long distances, but if disorder is  fully introduced, 
the glide motions of slips in particular 
directions should be much limited. 
 In MD simulations of two component glasses, for example, 
such degree of disorder should be sensitive to   
the size-ratio of the two species.  In fact, 
it  was rather close to 1 in 
the simulation by  Deng {\it et al.} 
\cite{Argon_review,Deng}, 
where   {\it nano-crystalline order}
 was realized  \cite{Argon_c} because of rather weak frustration  and long  
slips with atomic thickness emerged  along 
the crystal axes. It is therefore informative 
to perform MD  simulations 
with various size-ratios and examine 
the  shapes of  local configuration changes.

(v) 
As a special ingredient of our theory, 
 our elastic energy is almost isotropic if the 
distance from the  center of 
a  unit cell in the $e_3$-$e_2$ plane is 
shorter than $0.5$ (see Figs.1-3). 
The snapshot at $\gamma=0.4$ in Fig.11 
demonstrates that 
most spatial points are in the mechanically stable 
regions in plastic flow  and hence are in the isotropic elasticity regions. 
Therefore, our results  such as 
the  orientations of well-developed shear bands 
should  be applicable to those in  
amorphous materials (which are isotropic on large 
scales).

(vi)  The dimensionless parameter   $\lambda$ in 
(2.23) is related to Poisson's ratio $\nu$ 
as in (3.2) and the dislocation energy 
depends on $\nu$. Our choice  $K_0/\mu_0=\lambda=1$ 
or $\nu=0$ is rather  unusual, since 
$K_0$ is usually considerably 
larger than $\mu_0$ in high density systems. 
The appropriateness of the 
 other choices   $\eta_0^*=1$ in (2.26) 
and $\epsilon_{\rm th}=0.1$ in Section 4 
should also be examined in future.

(vii) As another aspect, 
we mention the effect of 
elastic interaction  among   dislocations. 
In our case slips are more easily 
created around preexisting ones, as already 
reported in Ref.\cite{Bulatov}. 
In the insets of Figs.15 and 21 we can see a tendency 
of aggregation of dislocation cores. 
In real 3D crystals   
 a tangle of dislocations often 
appears as   a characteristic feature  
of fatigued states \cite{dislocation} and 
has also been realized in 3D large-scale 
simulations \cite{Kubin,Ab}. 
In the  literature the inter-dislocation
elastic interaction  is believed to yield 
 mesoscopic patterns in the dislocation 
distribution 
\cite{Kubin,Ab,dislocation}.

(viii) As already stressed in the introduction, 
in order to describe glass dynamics in a more satisfactory level, 
 we should try to  include 
a variable  representing the local free-volume  \cite{OnukiJ} 
and   the configurational frustration 
effect induced by the size 
disparity between the two species. 
This generalization should 
be essential  with increasing $T$ 
towards the glass transition. 
For example, we may predict 
a gradual diffusional increase of 
the free volume around dislocation cores 
\cite{OnukiJ},  which induces   configuration changes and 
presumably  resulting in significant aging effects. 
(The critical strains  $\gamma_{{\rm c}1}$ and $\gamma_{{\rm c}2} $ 
discussed below (3.23) and indicated 
in Fig.6 by arrows are  sensitively decreased 
in magnitude   by a small amount of 
the local free volume  near the dislocation cores.) 
In the present study, as shown in  Fig.22,  we have found 
no appreciable aging effect.

\section*{ACKNOWLEDGMENTS}

This work is supported by Grants in Aid for Scientific 
Research from the Ministry of Education, Science and Culture.
The author thanks  R. Yamamoto for valuable 
discussions and A. Furukawa for his help 
in writing the figures.  Thanks are also due to A.S. Argon 
for sending his papers and informative correspondence.

\vspace{5mm}
{\bf APPENDIX}
\setcounter{equation}{0}
\renewcommand{\theequation}{A.\arabic{equation}}
\vspace{5mm}

Here we examine the linearized equations 
of (2.16) and (2.17) around a homogeneously 
strained state with $\av{e_1}= 0$, 
$\av{e_3}= \gamma $,  and 
$\av{e_2}= \epsilon$.   We neglect the random stress 
and assume that all the deviations depend 
on space and time as $\exp(i{\bi k}\cdot{\bi r}+i \omega t)$. 
Then the deviation of the displacement vector 
$(\delta u_x,\delta u_y)$ obeys 
\bea
\Omega 
\delta u_x &=& 
 C_{xx}\delta u_x + C_{xy}\delta u_y , 
\nonumber\\
\Omega 
\delta u_y &=& C_{xy}\delta u_x + C_{yy}\delta u_y, 
\label{eq:A.1}
\ena
where 
\be 
\Omega=\rho \omega^2/k^2 - i\eta_0 \omega.
\label{eq:A.2}
\en   
The frequency $\omega$ is expressed in terms of $\Omega$ 
at small $k$ as 
\be 
\omega= \pm (\Omega/\rho)^{1/2} k + i({\eta}_0/{2\rho})k^2+ O(k^3).  
\label{eq:A.3}
\en 
The first term is the oscillation  
frequency for $\Omega>0$ 
and the second term viscous damping. 
The coefficients $C_{\alpha\beta}$
are expressed in terms of the 
coefficients $\Phi_{\alpha\beta}$ in (2.7) 
as 
\bea 
C_{xx}= (K_0+\Phi_{22})n_x^2+ 2\Phi_{23}n_xn_y + \Phi_{33}n_y^2, 
\nonumber\\
C_{yy}= (K_0+\Phi_{22})n_y^2- 2\Phi_{23}n_xn_y + \Phi_{33}n_x^2, 
\nonumber\\
C_{xy}= (K_0-\Phi_{22}+\Phi_{33})n_xn_y 
+ \Phi_{23}(n_x^2-n_y^2). 
\label{eq:A.4}
\ena 
Here  ${\bi n}=k^{-1}{\bi k}$ 
is the unit vector representing the direction of the 
wave vector $\bi k$. 
From (A.1) it follows the relation, 
\be  
(\Omega-C_{xx})(\Omega-C_{yy})=C_{xy}^2.
\label{eq:A.5}
\en

Let  the angle of $\bi n$ be 
$\varphi+\pi/2$ with respect to the $x$ axis; then,    
\be 
n_x= -\sin \varphi,\quad  
n_y= \cos \varphi. 
\label{eq:A.6}
\en 
After some calculations (A.5) is solved to give  
\bea 
\Omega &=& 
\frac{1}{2}(K_0+\Phi_{22}+\Phi_{33}) \nonumber\\
&&\hspace{-0.5cm}\pm \sqrt{\frac{1}{4}K_0^2+ A^2+B^2+ K_0(A\cos 4\varphi 
+B\sin 4\varphi)}, 
\label{eq:A.7}
\ena
where  $A=(\Phi_{22}-\Phi_{33})/2$ 
and $B= \Phi_{23}$.  As a function of $\varphi$ 
 the slowest mode is obtained if the combination 
$A\cos 4\varphi +B\sin 4\varphi$ 
takes the maximum 
$(A^2+B^2)^{1/2}$. The corresponding minimum 
of $\Omega$ is  given by  
\be 
\Omega_{\rm min}  =   
\frac{1}{2}(\Phi_{22}+\Phi_{33}) 
-\sqrt{A^2+ B^2} .
\label{eq:A.8}
\en  
Notice that $\Omega_{\rm min}$ is the smaller 
of the two eigenvalues of the matrix $\{\Phi_{\alpha\beta}\}$. 
We may draw two conclusions. 
(i) As the instability 
point is approached,  $\Omega_{\rm min}$ tends to zero 
with softening 
of the sound speed $(\Omega_{\rm min}/\rho)^{1/2}$ 
of the corresponding acoustic  mode. 
(ii) If the approximate expression 
 (2.8) is used, we have 
$A= 2G'' (\epsilon^2-\gamma^2)$ and 
$B= 4G'' \epsilon\gamma$ so that 
\be 
A\cos 4\varphi +B\sin 4\varphi= 
C\cos (4\varphi+ 2\alpha), 
\label{eq:A.9}
\en 
where $C=2|G''| (\epsilon^2+\gamma^2)$ and $\alpha$ is defined by (3.15). 
Thus the minimum condition for $\Omega$ yielding (A.8) 
is given by $4\varphi+ 2\alpha=2n\pi$ 
and is equivalent to (3.14). 
That is, 
 the slowest  mode, 
which  undergoes softening 
at the instability point,  has a wave vector   
perpendicular to the favorable slip 
orientations given by (3.14).  This is the case 
even far below the instability point.  
In addition, for the slowest mode, 
(A.1) gives 
\be 
{\delta u_y}/{\delta u_x}= ({\Omega_{\rm min}-C_{xx}})/{C_{xy} } 
= \cot\varphi, 
\label{eq:A.10}
\en 
if $\alpha$ is eliminated using  (3.14).  
Thus the deviation $\delta {\bi u}$ is perpendicular 
to the wavevector $\bi k$ or the slowest mode is a transverse sound.

\end{multicols}

\newpage
\begin{figure}[t]
\epsfxsize=3.8in
\centerline{}
\caption{\protect
The scaling  function  
${\Phi}(e_3,e_2)$ in  (2.5), 
which is   the  shear deformation energy density 
divided by $\mu_0$ with 
 one of the crystal axes being along the $x$ axis.
}
\label{pp}
\end{figure}
\begin{figure}[t]
\epsfxsize=3.4in
\centerline{}
\caption{\protect
 ${\Phi}(e_3,e_2)$ for various 
directions in the $e_3-e_2$ plane.  It  demonstrates 
isotropic behavior  for $e=(e_3^2+e_2^2)^{1/2}  < 0.5$. 
The curve for uniaxial stretching ($e_2=e$ and $e_3=0$) 
coincides with  the curve of $n=5$ in the figure.  
The instability points are marked, 
which separate stable and unstable regions
}
\label{fsqt}
\end{figure}

\begin{figure}[t]
\epsfxsize=2 in
\centerline{}
\caption{\protect
The stable regions of our elastic energy 
are shown in white, where the two eigenvalues 
of $\{ \Phi_{\alpha\beta}\}$ are both positive. 
In the gray region one of them  
is negative, and  in the black regions 
both of them are  negative. 
}
\label{pp}
\end{figure}


\begin{figure}[b]
\epsfxsize=3.8in
\centerline{}
\caption{\protect
 The $x$ component of the displacement for a 
numerically calculated type C slip with length 20 
along the $x$ axis in the unstrained 
condition. }
\label{fsqt}
\end{figure}


\begin{figure}[b]
\epsfxsize=3.9in
\centerline{}
\caption{\protect
The displacement vector $\bi u$ around 
type C and type CC slips with length 10. 
The orientations in (a) are most favorable 
in shear deformation, 
while those in (b) are 
most favorable in uniaxial stretching. 
The arrows are from the original 
undeformed  position 
to the displaced position.  }
\label{fsqt}
\end{figure}


\begin{figure}[t]
\epsfxsize=3.8 in
\centerline{}
\caption{\protect
 The slip energy $F_{\rm slip} $ 
vs applied shear strain $\gamma$ 
for type C slips with length $\ell=10, 20,$ and $30$ 
oriented along the $x$ axis.  For $|\gamma| < 0.05$ 
the relation (3.12) holds with slope $-\ell$. 
 The arrows indicate 
 the instability points 
where expansion or shrinking of the slips occur.  }
\label{fsqt}
\end{figure}


\begin{figure}[t]
\epsfxsize=7in
\centerline{}
\caption{\protect
  The slip energy density around a type C slip with length 20 
for $\gamma=0$ in (a), 0.065 in (b), and -0.04 in (c). 
In the middle region between the dislocations 
at the ends,  the elastic energy 
 is decreased in (b) and increased in (c). }
\label{fsqt}
\end{figure}


\begin{figure}[t]
\epsfxsize=4.5in
\centerline{}
\caption{\protect
Coincidence of 
the derivative $\p F_{\rm slip} /\p \gamma$ ($\times$) 
and  the space integral of 
$\sigma_{xy}-\gamma$ ($+$)  confirming the general relation 
(3.17).  These   
quantities are calculated  from the steady slip solutions.
  }
\label{fsqt}
\end{figure}


\begin{figure}[b]
\epsfxsize=4.2in
\centerline{}
\caption{\protect
  }
\label{fsqt}
\end{figure}

 FIG.9.~ 
The slip energy difference 
$\Delta F_{\rm slip}(\ell) 
= F_{\rm slip}(\ell)- F_{\rm slip}(20)$ 
in the range $20\le \ell \le 21$ 
obtained for the extrapolated displacement (3.22) for $\gamma=-0.04, 0, 
0.08$, and $0.1$. 
The position of the maximum is a decreasing 
function of $\gamma$. The maximum position approaches 
$\ell=21$ as $\gamma \rightarrow \gamma_{{\rm c}1}  \cong -0.09$ 
and $\ell=20$ as $\gamma \rightarrow \gamma_{{\rm c}2}  \cong 0.12$. 
The inset shows the displacement vector 
${\bi u}_{21}-{\bi u}_{20}$ needed for 
slip growth by unit length from $\ell=20$ to 21.


\begin{figure}[t]
\epsfxsize=5.2 in
\centerline{}
\caption{\protect
The stress-strain curves in shear flow.  
The initial states are 
defectless for   $\gdot=10^{-4}$ and  $10^{-3}$ (with the 
higher peak). Four slips are initially prepared 
for  $\gdot=10^{-3}$ (with the lower  peak). 
Embedded  are snapshots of $\delta e_3=e_3-\gamma$ 
at  $\gamma=0.3$  at the inception of slip formation 
and at $\gamma=0.4$ 
in the plastic flow regime.
  }
\label{fsqt}
\end{figure}

\begin{figure}[b]
\epsfxsize=4in
\centerline{}
\caption{\protect
  Mechanically stable  regions 
(white) and unstable points (black) 
after application of 
shear flow  with $\gdot=10^{-3}$. The initial state is 
crystalline without defects. 
At the stable points  the eigenvalues 
of the matix $\{\Phi_{\alpha\beta}\}$ defined below (2.7) 
are both positive, while at the unstable points   
at least one of them is negative. 
 At $\gamma=0.29$ the system is close to the peak position, 
at $\gamma=0.30$ slips are appearing, 
and at $\gamma=0.4$ the unstable points are fluctuating in time 
near the dislocation 
cores.}
\label{fsqt}
\end{figure}

\begin{figure}[t]
\epsfxsize=4.0in
\centerline{}
\caption{\protect
The displacement deviation
 $\delta{\bi u}= {\bi u}- (\gamma y,0)$ 
in the plastic flow regime 
under shear strain at $\gamma=0.4$ with $\gdot=10^{-4}$. 
A  $1/4$ region ($64\times 64)$  of the total system 
is shown. The arrows are from the original position 
at $t=0$ in a perfect crystal 
 to the displaced position in  plastic flow. 
  }
\label{fsqt}
\end{figure}

\begin{figure}[t]
\epsfxsize=6.0in
\centerline{}
\caption{\protect
  Time-evolution  of $\delta e_3$ and $\delta f_{\rm el} $ 
in shear flow when   four slips are placed  in the initial state.  }
\label{fsqt}
\end{figure}

\begin{figure}[t]
\epsfxsize=3.8in
\centerline{}
\caption{\protect
   }
\label{fsqt}
\end{figure}
FIG.14.~
The stress-strain curve for cyclic shear deformation 
in the  first two cycles at $\gdot=\pm 10^{-3}$ 
with period $10^3$.  Once  $\gamma \gs 0.3$ in the first 
cycle,  high-density 
dislocations are created.  The shear stress vanishes at the three 
points A, B, and C ($\times)$.
\begin{figure}[t]
\epsfxsize=4.2in
\centerline{}
\caption{\protect
  The average 
elastic energy density $\av{f_{\rm el}}$ vs 
the strain $\gamma(t)$ for cyclic shear deformation in the first 
two cycles.    At the  points A, B, and C 
the shear stress vanishes. In the inset 
the snapshot of $f_{\rm el}$ at the point A 
is shown, where 
$\av{\sigma_{xy}}=  
\gamma_{\rm el}=0$ and the black points represent 
dislocation cores.   }
\label{fsqt}
\end{figure}
\begin{figure}[t]
\epsfxsize=4in
\centerline{}
\caption{\protect
   The  energy density $f_{\rm D}$ defined by (4.2) 
vs the strain $\gamma(t)$ 
for cyclic shear deformation in the first 
two cycles. It is  the defect  energy 
density in plastic flow, but in the preplastic regime  
it arises from the heterogeneities in the strains 
and is enhanced at the onset of plastic flow. }
\label{fsqt}
\end{figure}


\begin{figure}[t]
\epsfxsize=5.2in
\centerline{}
\caption{\protect
  The stress-strain curves under uniaxial stretching.  
The initial states are 
defectless for   $\epsilon=10^{-4}$ and  $10^{-3}$ (with the 
higher peak). Four slips are initially prepared 
for  $\epsilon=10^{-3}$ (with the lower  peak). 
Embedded are snapshots of $\delta e_2=e_2-\epsilon$ 
at  $\epsilon=0.31$  at the inception of slip formation 
and  at $\epsilon=0.46$ in the plastic flow regime.}
\label{fsqt}
\end{figure}

\begin{figure}[t]
\epsfxsize=4in
\centerline{}
\caption{\protect
   The displacement deviation 
$\delta{\bi u}={\bi u}- (\epsilon x/2,-\epsilon y/2)$ 
in the plastic flow regime 
under uniaxial stretching 
 at $\epsilon=0.46$ with $\epsilon=10^{-3}$. 
A  $1/4$ region ($64\times 64)$  of the total system 
is shown.  The arrows are from the original position 
at $t=0$ in a perfect crystal 
to the displaced position in plastic flow.}
\label{fsqt}
\end{figure}

\begin{figure}[t]
\epsfxsize=3.8in
\centerline{}
\caption{\protect
Time-evolution  of $ e_3$, $\delta e_2$, and $\delta f_{\rm el} $
under uniaxial stretching when  
four slips are placed  at $t=0$.  
The initial orientations are marginal and 
new slips with the favorable orientations 
appear from the ends  of the initial  slips. 
  }
\label{fsqt}
\end{figure}

\begin{figure}[t]
\epsfxsize=3.2in
\centerline{}
\caption{\protect
 }
\label{fsqt}
\end{figure}
FIG.20. 
The stress-strain curve for cyclic uniaxial stretching  
in the  first four cycles  at $\dot{\epsilon} =\pm 10^{-3}$ 
with period $10^3$.   Once  $\epsilon \gs 0.3$ in the first 
cycle,  high-density dislocations are created.  In this case 
the stress-strain relation becomes nearly periodic from the second cycle.  

\begin{figure}[t]
\epsfxsize=4.2in
\centerline{}
\caption{\protect
  The average total 
elastic energy density $\av{f_{\rm el}}$ (solid line) 
and the defect part 
$f_{\rm D}$ (broken line) vs the strain $\epsilon(t)$ 
for cyclic uniaxial stretching in the first 
two cycles.     In the inset 
the snapshot of $f_{\rm el}$ at the point 
A in the first cycle in  Fig.20 is shown,  where 
$\av{\sigma_{xx}-\sigma_{yy}}=  
\epsilon_{\rm el}=0$ and the black points represent 
dislocation cores. }
\label{fsqt}
\end{figure}

\begin{figure}[t]
\epsfxsize=4.4in
\centerline{}
\caption{\protect
Relaxation of the average energy density $\av{f_{\rm el}}(t)$ 
for $\epsilon_{\rm th}=0.1$ (a) (upper curve) 
and $\epsilon_{\rm th}=0.25$ (b) (lower curve).
The shear flow is stopped at time $t_{\rm A}= 600$ 
(at the point A in Fig.14) and 
the system is relaxed  for a time period of $t_{\rm w}= 10^5$. 
The  energy decreases are extremely small 
as compared to the initial values, so there are almost no 
appreciable aging behavior here.   
The inset displays the stress-strain curves, 
$\av{\sigma_{xy}}$ vs $\Delta \gamma(t)$,
after the shear is switched on again 
for  $t \ge t_{\rm A}+t_{\rm w}$, where the solid line corresponds to 
(a) and the noisy broken line to (b). 
The  $\Delta \gamma(t)= \dot{\gamma}(
t-t_{\rm A}-t_{\rm w})$ is the excess strain with $\gdot=10^{-3}$. 
Almost identical stress-strain 
 curves are obtained for any waiting time $t_{\rm w}$ 
shorter than $10^5$. 
  }
\label{fsqt}
\end{figure}


\begin{references}
\bibitem{uni} Typical   applied stress 
is tensile, compressive, 
or shear.   In experiments uniaxial 
deformation is  more usual than shear deformation.
\bibitem{commentp} If (2.12) is assumed, the elastic energy 
on the basis of (2.5) becomes a  
functional of $\bi u$ only. Then the resultant stress tensor 
following from (2.13)  contains a small anti-symmetric 
part or $\sigma_{xy}-\sigma_{yx}\neq 0 $  
on the order of $e^6$ from the expansion (2.6). This suggests 
room of improvement of  our nonlinear 
strain theory.
\bibitem{commentV} In 2D the viscous 
stress tensor is written as $\eta_0(\nabla_i v_j + 
\nabla_jv_i- \delta_{ij}\nabla\cdot{\bi v}) 
+\zeta_0 \delta_{ij}\nabla\cdot{\bi v}$, 
where $\eta_0$ and $\zeta_0$ are  the shear and bulk 
viscosities, respectively \cite{Landau}.
\bibitem{commentc} If $\rho$ is included as a dynamic variable, 
$\rho$ obeys 
the continuity equation 
${\p}\rho/{\p t}  = -\nabla\cdot({\rho{\bi v}})$ 
and the right hand side of (2.17) should be replaced 
by $\rho [{\p}/{\p t}+  
{\bi v}\cdot\nabla  ] {\bi v}$. 
  Then the self-consistency relation (2.20) is kept to be valid. 
 For large strains of order 1 this generalization is needed.
\bibitem{energy} This slip energy may be 
calculated in the linear elasticity theory \cite{Landau}. 
If (3.3) is substituted 
into  the elastic energy expression, 
  a logarithmic space  integral ($\propto \int dr r^{-1})$ 
appears in the  range  
$a\ls r\ls \ell$.  The lower 
bound is provided by     nonlinear elasticity 
theory.  This slip energy also follows from 
 (3.11).
\bibitem{climb} In 2D, the glide direction 
is along the Burgers vector $\bi b$, 
and the climb direction is perpendicular to 
$\bi b$ \cite{Friedel,Nabarro}.
\bibitem{crystal} 
The Peierls potential  energy  
 depends on the ratio of 
the atomic spacing across the  glide 
plane and the magnitude of the Burgers vector 
\cite{Friedel,Nabarro,P,N}. 
As a result,  the critical stress 
causing dislocation motion (Peierls stress) can  
strongly depend on the orientation of the glide plane   
with respect to the crystal axes.
\end{references}
\end{document}